# Automatic EEG Independent Component Classification Using ICLabel in Python

Arnaud Delorme
*CerCo CNRS, Paul Sabatier University, Toulouse, France*
*SCCN, INC, UCSD, La Jolla CA, USA*
https://orcid.org/0000-0002-0799-3557

Dung Truong
*SCCN, INC, UCSD, La Jolla CA, USA*
dutruong@ucsd.edu

Luca Pion-Tonachini
*SCCN, INC, UCSD, La Jolla CA, USA*
lucapton@gmail.com
https://orcid.org/0000-0002-4018-7776

Scott Makeig
*SCCN, INC, UCSD, La Jolla*
https://orcid.org/0000-0002-9048-8438

*Abstract*—ICLabel is an important plug-in function in EEGLAB, the most widely used software for EEG data processing. A powerful approach to automated processing of EEG data involves decomposing the data by Independent Component Analysis (ICA) and then classifying the resulting independent components (ICs) using ICLabel. While EEGLAB pipelines support high-performance computing (HPC) platforms running the open-source Octave interpreter, the ICLabel plug-in is incompatible with Octave because of its specialized neural network architecture. To enhance cross-platform compatibility, we developed a Python version of ICLabel that uses standard EEGLAB data structures. We compared ICLabel MATLAB and Python implementations to data from 14 subjects. ICLabel returns the likelihood of classification in 7 classes of components for each ICA component. The returned IC classifications were virtually identical between Python and MATLAB, with differences in classification percentage below 0.001%.

## I. Introduction

EEGLAB is a powerful and widely utilized open-source scripting environment running on MATLAB (The Mathworks, Inc.) to support the analysis of electroencephalographic (EEG) data [1]. EEGLAB provides a comprehensive suite of tools for processing of single-trial and trial-averaged EEG data across experimental conditions, including data import, visualization, preprocessing (including artifact rejection and filtering), and modern analysis techniques, including independent component analysis (ICA) [1,2]. The toolbox has gained significant use in the neuroscience community [4] through its user-friendly interface and extensive functionalities that facilitate the exploration of complex EEG dynamics [3].

A notable extension of EEGLAB is the ICLabel plug-in of Pion-Tonachini et al. [5], which employs a machine learning approach to classify independent components of EEG data (ICs) into seven source-type categories: Brain, Eyes, Muscle, Heart, Line Noise, Single Channel, and Other — thereby streamlining the process of identifying and separating non-brain signals in EEG data, allowing researchers to focus on relevant brain signals while minimizing the influences of non-brain ('artifact') sources [6].

The combination of EEGLAB-supported ICA decomposition and the ICLabel plug-in represents a significant advance in EEG research tools, one that enables researchers to efficiently preprocess and analyze large or small datasets. However, using EEGLAB for automated processing on high-performance computing (HPC) platforms poses challenges. Automated workflows involving ICA decomposition and IC classification using the ICLabel plug-in run only in the MATLAB environment. This dependency can limit access to HPC resources, as MATLAB licenses may not always be available. While Octave [11] provides an open-source alternative capable of running most EEGLAB processing pipelines on HPC systems, it cannot execute the ICLabel plug-in because of the specialized neural network architecture used in the ICLabel classifier. Consequently, there is a need for a solution that allows seamless integration of ICLabel-based data processing workflow on HPC platforms that do not support MATLAB.

Although a derivative of ICLabel for Python does exist [7], it has not undergone formal validation and lacks a peer-reviewed reference article to support its accuracy and reliability. Furthermore, it uses a distinct computational approach from the published ICLabel implementation. These differences may affect the consistency of classification outcomes compared to the MATLAB-based function, raising concerns about the tool's validity for scientific research. As a result, there remains a critical need for a Python-coded ICLabel extension that (1) faithfully reproduces the original function's behavior, (2) has

Funded by NIH R01NS047293.

been rigorously validated, and (3) is supported by a reference publication. Here, we report a Python-coded function that meets this need.

## II. METHODS

### A. Code conversion

Our strategy for code conversion from MATLAB to Python (see Fig. 1) used an AI-assisted approach combined with manual debugging to ensure accuracy. The reason for using AI was to speed up the conversion process. Initially, the original MATLAB code (in Fig. 1, the dummy function *add.m*), which defines a simple function, and a script that calls this function (*call_add.m*) are subjected to AI-based code translation (in our case, using ChatGPT 4o). This process converts the MATLAB function into an equivalent Python function (*add.py*) and a Python calling script (*call_add.py*) that includes necessary libraries like *scipy.io* [12].

After conversion, we compared the results of running the original MATLAB function and the Python-translated function on actual EEG data. This comparison involved first executing the original ICLabel MATLAB code, next calling the Python equivalent from within MATLAB using the MATLAB *eval* function, and finally loading its output back into MATLAB. If the percentage difference (see Methods section C) between the MATLAB and Python version results was more than the minimal threshold of 0.1%, we used manual debugging to resolve the discrepancies. This iterative process ensured that the translated code functions equivalently to the original.

### B. Test datasets

To test the code conversion, we used the EEGLAB tutorial dataset *eeglab_data_epochs_ica.set* [1], a 32-channel dataset that includes an ICA decomposition performed using Infomax ICA [9]. Although this ICA decomposition was performed before the development of ICLabel, the algorithm used for fitting ICA in ICLabel aligns with the one applied to the components in *eeglab_data_epochs_ica.set*.

We have also compared the ICLabel versions using the EEGLAB tutorial study (STERN study) [10]. This dataset comprises 13 subjects and precomputed ICA weights using the *runica.m* function of EEGLAB. Each subject is comprised of 3 datasets for different experimental conditions (Memory, Ignore, Probe). For comparison between MATLAB and Python, we only used the Memory condition for each subject.

These datasets were selected for their accessibility and the fact that they contain precomputed ICA weights, allowing users to easily replicate our results.

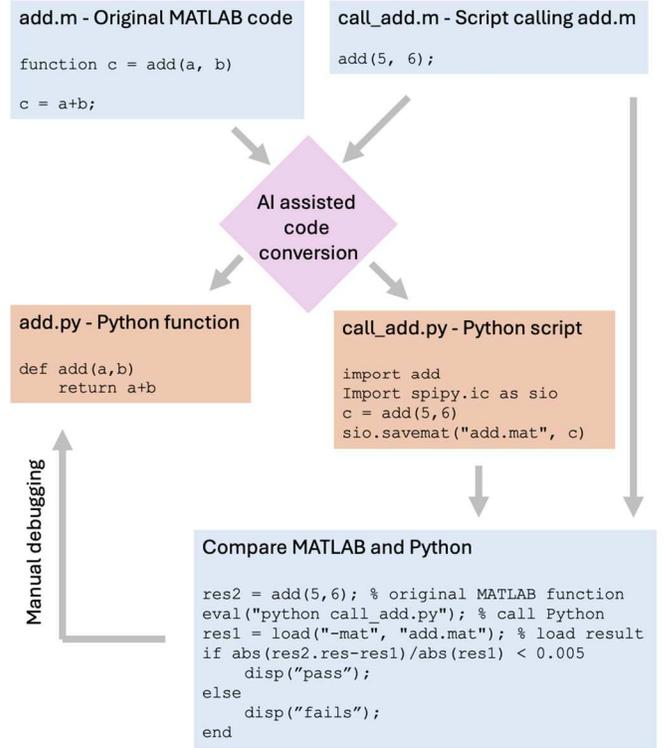

Figure 1. Workflow for AI-assisted conversion of MATLAB code to Python code. Here, an example *add.m* MATLAB function is converted into a Python function, *add.py*. A script (*call_add.m*) to call the MATLAB function is also converted into an equivalent Python script (*call_add.py*). The converted Python script saves the result to a *.mat* file for compatibility with MATLAB. The comparison step involves running both the original MATLAB script and the converted Python script, loading the results in MATLAB, and checking whether the outputs match within an acceptable tolerance (see Methods). Manual debugging is used to address any discrepancies.

### C. Assessing output differences between the MATLAB and Python implementations

To assess the difference between the MATLAB and Python results for a given feature (scalp topography, power spectrum, autocorrelation), we computed the maximum percentage absolute difference $mdp$.

$$mdp = \max_i \left( abs \left( \frac{f_i^{MATLAB} - f_i^{Python}}{f_i^{MATLAB}} \right) \right)$$

Where $f_i^{MATLAB}$ is a feature value in MATLAB and $f_i^{Python}$ is the corresponding feature value in Python. $i$ indexes the values for a given feature across subjects, components, scalp channels (for scalp topographies), and frequencies (for power spectra and autocorrelations).

## III. RESULTS

### A. IC scalp topographies

The first information ICLabel relies on are components' scalp topography computed using the MATLAB function *topoplotFast.m*, adapted from the standard EEGLAB function *topoplot.m*. This function interpolates component scalp maps (using average reference) on a 32x32 scalp grid. Because interpolation methods vary between scientific libraries of programming languages, this was the most challenging method to implement. For scalp interpolations, MATLAB uses the "v4" method (MATLAB so-called version-4 algorithm), which uses biharmonic spline interpolation–a smooth, scattered data interpolation technique that fits a surface through given data points while minimizing the bending energy of the surface.

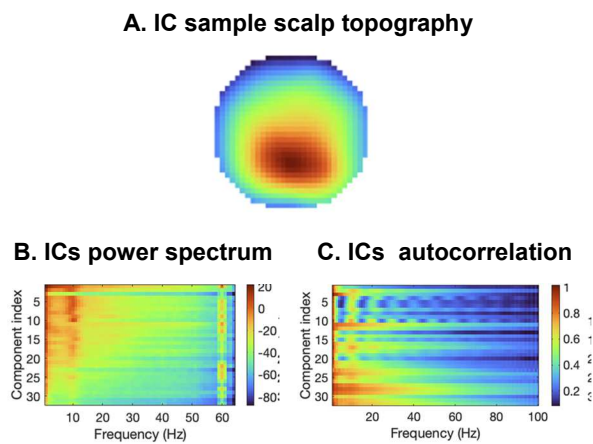

Figure 2. A. Sample IC scalp topography in ICLabel (the first component of EEGLAB test dataset–see Methods). B. Power spectral density for 32 components for the EEGLAB test dataset. C. Spectral autocorrelation values for the 32 ICs of the EEGLAB test dataset in the MATLAB and Python versions.

We were able to re-implement this MATLAB scalp map interpolation function in Python (Fig. 2A). For the EEGLAB tutorial dataset (see Methods), the mean difference in interpolated scalp map values across all components was smaller than $10^{-5}$ percent. The maximum absolute difference across all components of all datasets for the EEGLAB tutorial study was also below $10^{-5}$ percent (Fig. 3A) indicating a good match between MATLAB and Python implementations.

### B. Power spectrum density

ICLabel computes log-power spectral density using its *eeg_ersp.m* function (Fig. 2B). For the EEGLAB tutorial dataset, the maximum absolute difference between MATLAB and Python across all components was $10^{-3}$ percent. The maximum absolute difference for the EEGLAB tutorial study across all components and all subjects was $10^{-3}$ percent (Fig. 3C) indicating a good match between MATLAB and Python implementations.

### C. Autocorrelation function

ICLabel computes component activity autocorrelations using three functions (*eeg_autocorr*, *eeg_autocorr_welch*, and *eeg_autocorr_fftw*) (Fig. 2C). The function used depends on the type of data. For continuous data, 3-sec data epochs are extracted. If there are more than five epochs, *eeg_autocorr_welch* is used; otherwise, *eeg_autocorr* is used. For data consisting of a series of extracted data epochs, ICLabel uses *eeg_autocorr_fftw*. All functions are similar in their implementation to compute autocorrelation, with the *eeg_autocorr_fftw* bypassing the window extraction step and the *eeg_autocorr* using a time-based method implemented as *eeg_autocorr_welch*.

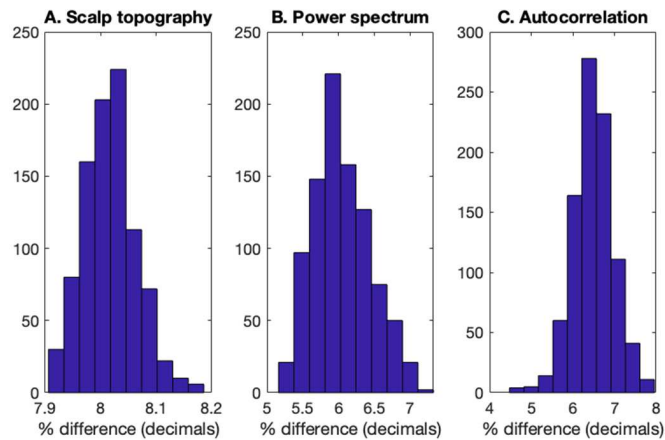

Figure 3. Maximum likelihood difference histograms (across the 13 subjects (897 ICA components) tutorial study–see Methods) between the MATLAB and Python values for the three measures shown in Fig. 2. The abscissa shows the number of decimals. For example, 4 corresponds to 0.0001. The ordinate shows the number of components in each bin.

For the EEGLAB tutorial dataset, the maximum absolute autocorrelation difference between the MATLAB and Python functions across all components was 0.01%. The maximum absolute difference for the tutorial study across all components and all subjects was also 0.01% (Fig. 3C).

### D. Neural network implementation

We implemented the default method of the EEGLAB ICLabel plug-in using Pytorch. This is the method most users use. In ICLabel, a neural network is trained to classify ICs as belonging to seven data source categories [5] based on their characteristics (Fig. 2). The network uses input features computed in previous sections (scalp maps and IC activity power spectral densities and auto-correlations). The ICLabel model outputs probabilistic estimates for 7 component classes (Brain, Eyes, Muscle, Heart, Line Noise, Single Channel, and Other) indicating the likelihood that a given IC belongs to each class. This probabilistic approach helps handle ambiguities and allows a more accurate descriptive classification across datasets and recording conditions.

| Name | Type | Input | N-out | Kernel | Stride | Padding |
|---|---|---|---|---|---|---|
| Topo-1 | Conv | *Topo data* | 128 | 4x4 | 2 | 1 |
| Topo-2 | Conv | Topo-1 | 256 | 4x4 | 2 | 1 |
| Topo-3 | Conv | Topo-2 | 512 | 4x4 | 2 | 1 |
| PSD-1 | Conv | *PSD data* | 128 | 1x3 | 1 | (0,1) |
| PSD-2 | Conv | PSD-1 | 256 | 1x3 | 1 | (0,1) |
| PSD-3 | Conv | PSD-2 | 1 | 1x3 | 1 | (0,1) |
| ACF-1 | Conv | *ACF data* | 128 | 1x3 | 1 | (0,1) |
| ACF-2 | Conv | ACF-1 | 256 | 1x3 | 1 | (0,1) |
| ACF-3 | Conv | ACF-2 | 1 | 1x3 | 1 | (0,1) |
| R1 | Reshape | ACF-3 | (100, 1, 1) | | | |
| R2 | Reshape | PSD-3 | (100, 1, 1) | | | |
| C1 | Concat. | R2 | dim=2 | | | |
| C2 | Concat. | C1 | dim=3 | | | |
| C3 | Concat. | R1 | dim=2 | | | |
| C4 | Concat. | C3 | dim=3 | | | |
| C5 | Concat. | Topo-3, C2, C4 | | | | |
| Discr | Conv | C5 | 7 | 4 | 1 | 0 |
| Final | Softmax | | | | | |

Table 1. ICLabel neural network architecture. The term "Topo" refers to "scalp topography," while PSD stands for power spectral density, ACF stands for "autocorrelation function" and LReLU for "leaky ReLU" (with a leakage factor of 0.2).

When we compared the MATLAB and Python neural network implementations using the same MATLAB-precomputed features as input for both the tutorial dataset and the tutorial study, the maximum difference was $10^{-5}$ percent.

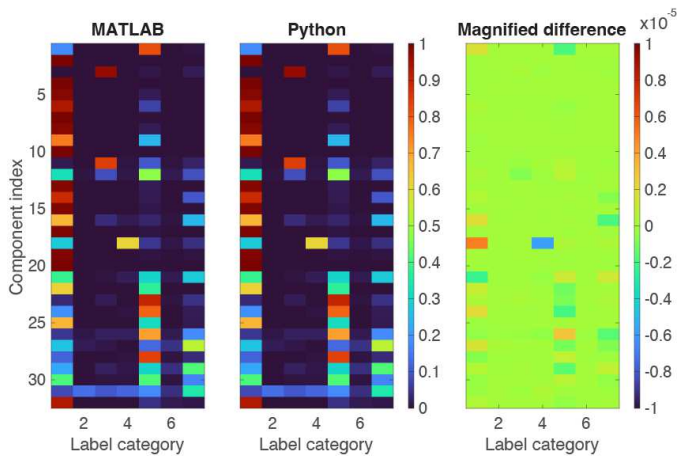

Figure 4. Results of the ICLabel pipeline in MATLAB and in Python for the EEGLAB tutorial dataset, and their difference magnified between -0.001% and 0.001% For practical applications, the MATLAB and Python implementations provide identical results.

Upon executing the complete MATLAB and Python pipeline, we observed, for both the tutorial dataset and the 13 subjects in the tutorial study, a maximum difference in IC likelihoods below 0.001%. In this test, the entire ICLabel process was performed either in Python or MATLAB. The minuscule difference for one dataset (the first dataset of the tutorial EEGLAB study) is shown in Fig. 4.

The ICLabel plugin classifies independent components into seven categories. One approach involves selecting the class (Brain, Eyes, Muscle, Heart, Line Noise, Single Channel, and Other) with the maximum likelihood for each component. In the tutorial dataset and the tutorial study (see Methods), when comparing Python with MATLAB ground truth, no components were misclassified when assigned to the class with the highest likelihood among the seven classes.

*E. Usage*

ICLabel is used in Python in the same way it is used in MATLAB. In Fig. 5, we show how to import the *eeglab* Python library, apply ICLabel, and display the class probabilities for each ICA component of the EEGLAB tutorial dataset. This code may be used on most high-performance computing (HPC) platforms where MATLAB licenses are not available.

```python
from eeglab import pop_loadset
from eeglab import iclabel

# load EEGLAB dataset
EEG = pop_loadset('eeglab_data_epochs_ica.set')

# apply IClabel
EEG = iclabel(EEG)

# Show component classes
ic_classes = EEG['etc']['ic_classification']\
['ICLabel']['classifications']
print(ic_classes)
```

Figure 5. Python script to call ICLabel on an EEGLAB dataset.

MNE is the most widely used Python package for processing EEG data [8]. It is also possible to apply the Python version of ICLabel to an MNE dataset. The *eeg_mne2eeglab* Python function of the Python *eeglab* package allows the conversion of an MNE structure to a format that the *iclabel* function can use.

*Code availability.* The code for the Python package is available at *https://github.com/sccn/iclabel_python*.

## IV. DISCUSSION

The results across 14 datasets (the EEGLAB tutorial dataset plus 13 other datasets of the EEGLAB tutorial study) showed that the Python implementation of ICLabel closely matches that of the original MATLAB version, with differences in computed class likelihoods below 0.001%.

Automated workflows, particularly those combining ICA decomposition and component classification using tools such as ICLabel, can significantly streamline the process of non-brain artifact identification and separation, allowing researchers to process large datasets more efficiently. Integrating a Python-based alternative to ICLabel enhances the availability of the ICA/ICLabel approach in high-performance computing environments.

Developing a Python-compatible ICLabel implementation also represents a significant step toward fostering collaboration between the (now largely distinct) MATLAB and Python electrophysiological research communities. By enabling seamless cross-platform processing and providing a rigorously validated implementation of ICLabel in Python, this effort also bridges the gap between two widely used tool environments, EEGLAB and MNE, laying the groundwork for the creation of more robust, scalable, and comparable algorithms for EEG data analysis.

Looking forward, this integration opens avenues for further advancements in computational neuroscience, including the development of new machine-learning architectures and improved methods for ICA-based artifact rejection and data classification.